\documentstyle[12pt,twoside]{article}
\begin{document}
\begin{center}

{\Large\bf  DOES QUANTUM COSMOLOGY\\[5PT]
PREDICT A CONSTANT DILATONIC FIELD?\\[5PT]}
\medskip
 
{\bf F.G. Alvarenga\footnote{e-mail: flavio@cce.ufes.br},
A.B. Batista\footnote{e-mail: brasil@cce.ufes.br} and
J. C. Fabris\footnote{e-mail: fabris@cce.ufes.br}}  \medskip

Departamento de F\'{\i}sica, Universidade Federal do Esp\'{\i}rito Santo, 
CEP29060-900, Vit\'oria, Esp\'{\i}rito Santo, Brazil \medskip

\end{center}
 
\begin{abstract}

Quantum cosmology may permit to determine the initial conditions of the Universe.
In particular, it may select a specific model between many possible classical models.
In this work, we study a quantum cosmological model based on the string effective action
coupled to matter. The Schutz's formalism is employed in the description of the fluid.
A radiation fluid is considered.
In this way, a time coordinate may be identified and the Wheeler-DeWitt equation reduces
in the minisuperspace to a Schr\"odinger-like equation. It is shown that, under some
quite natural assumptions, the expectation values indicate a null axionic field and
a constant dilatonic field. At the same time the
scale factor exhibits a bounce revealing a singularity-free cosmological model. In some cases, the mininum value of the scale
factor can be related to the value of gravitational coupling.

\vspace{0.7cm}
PACS number(s): 98.80.Hw, 04.60.Gw
\end{abstract}

\section{Introduction}

Quantum cosmology is, in some sense, a theory of initial conditions \cite{halliwell,adm}. A classical model
depends on many initial data that must be fixed in order to have an agreement
with observations. There is a hope that the quantum cosmology may lead to specific predictions
concerning the values of at least some input parameters. One of the most promising candidates
for a unified theory of all interactions is the string theory
\cite{polchinsky,gasperini}. In string theory, the gravitational coupling is in principle a function of the space-time coordinates, being connected
with the expectation value of the dilatonic field. Moreover, the string action contains other
fields like the axion. The problem we treat here is the quantum scenario coming from the
string effective action in four dimensions. In particular, we are interested in which are
the predictions of quantum cosmology for the evolution of the dilatonic and axionic fields,
besides the scale factor.
\par
One of the most important problems in studying quantum cosmological models is the absence of
an explicit time coordinate \cite{isham,halliwell1}. There are many attempts to solve this special feature of the
quantum cosmological program. Here we adopt an interesting proposal: if ordinary matter is introduced in the model
through the Schutz's formalism \cite{schutz1,schutz2}, dynamical degrees of freedom are attributed to matter.
Quantizing the gravity system in presence of matter described by the Schutz's formalism,
we can obtain in the minisuperspace approach a Schr\"odinger-like equation, where the
conjugate momentum associated with matter field appears linearly. Hence, this matter field
may play the role of time. This approach has
been extensively employed in the literature, revealing in general the
suppression of the initial singularity \cite{rubakov,demaret,nivaldo1,nivaldo,brasil1,brasil2}.  Effectively, through this procedure, we end up with a genuine
Schr\"odinger equation, and all the machinery of ordinary quantum mechanics can be employed
in order to obtain specific predictions for the evolution of the Universe.
\par
Hence, we consider the string effective action in four dimensions in presence of matter.
More specifically, a radiative fluid is coupled to the string effective action. One of
the aims in considering a radiative fluid is to keep as far as possible in the
string context. There exists electromagnetic field terms
in the Ramond-Ramond sector of the string action. In this way, the inclusion of the radiative
fluid may not imply to get out of the pure string case. This radiative fluid is quantized
through the aid of the Schutz's variables, leading to a "time" variable. We parametrize
the string action by two parameters, $\omega$ and $n$. The parameter $\omega$ reflects the
dilatonic coupling. In the strict string case $\omega = - 1$. But we keep it as a free parameter
in order to consider more general frameworks, like brane models. The parameter $n$ is linked
with the coupling between the dilatonic and axionic field. Again in the strict string frame,
$n = - 1$. But, we also keep it arbitrary in order to include other configurations.
\par
Quantizing the string effective action in presence of a radiative fluid in the minisuperspace
we obtain a Schr\"odinger-like equation. We solve it for two distinct cases: $\omega > -3/2$,
the "normal" case, since it leads to a positive energy for the scalar field in the Einstein's
frame; $\omega < - 3/2$, the "anomalous" case, since the energy of the scalar field becomes
negative in the Einstein's frame. In general, the eigenfunctions are
not square integrable. For the normal case a superposition of them forming a wave packet can not
cure this pathologie, which is reflected in the infinite value of the norm of the wave
function. Hence, no quantum cosmological model can be constructed for this
case unless the dilatonic field
is fixed as a constant from the begining. This results in the traditional general relativity model \cite{nivaldo}.
\par
However, in the anomalous case it is possible to have square integrable wave functions. Wave
packets may be constructed. The expectation values for the scale factor, dilatonic field and
the axionic field may be evaluated. Surprisingly again, the expectation value for the dilatonic
field does not depend on time. At the same time, the expectation value for the axionic field
is zero. Hence, the anomalous quantum cosmological model coming from string theory coupled
to a radiative fluid predicts a constant gravitational coupling and no axionic field.
Morevoer, the scale factor exhibits a bounce. The minimum value for the scale factor can
be related to the value of the gravitational coupling. Inserting the known value of $G$,
we find that the minimum of the scale factor remains well above the Planck's length when
$\omega \rightarrow -3/2$; however, it becomes highly transplanckian when $\omega \rightarrow
- \infty$.
\par
In next section, we give details on the construction of the quantum system for the string effective
action with a radiative fluid described by the Schutz's formalism. In section $3$, we solve
the resulting Schr\"odinger-like equation and construct the corresponding wave packets.
In section $4$, the predictions for the evolution of the dynamical variables are presented.
In section $5$ we analyze the generality of the results taking into account the assumptions
made and the formalism employed, presenting our conclusions.

\section{The quantum model}

The Neveu-Schwartz sector of the string effective action is given by the Lagrangian density
\cite{copeland}
\begin{equation}
L = \sqrt{-g}e^{-\sigma}\biggr\{R + \sigma_{;\rho}\sigma^{;\rho} - e^{2\sigma}\chi_{;\rho}\chi^{;\rho}\biggl\} + L_m \quad ,
\end{equation}
where $\sigma$ is the dilatonic field, $\chi$ is the axionic field and $L_m$ is a term
taking into account the presence of ordinary matter. Through the redefinition
$\phi = e^{-\sigma}$, the Lagrangian takes the form
\begin{equation}
\label{Lag}
L = \sqrt{-g}\phi\biggr\{R - \omega\frac{\phi_{;\rho}\phi^{;\rho}}{\phi^2} - \phi^{n-1}\chi_{;\rho}\chi^{;\rho}\biggl\} + L_m \quad .
\end{equation}
Two free parameters, $\omega$ and $n$, have been introduced in order to take into account
more general frameworks besides the strict string effective action. The strict string case is characterized by
$\omega = - 1$ and $n = -1$. In this way, effective actions coming from $F$ theory,
supergravity theories or pure multidimensional models are included in the Lagrangian
(\ref{Lag}), as well as brane configurations \cite{duff}. In particular, some situations where $\omega$ can be largely negative are
taken into account.
\par
The Friedmann-Robertson-Walker metric is written as
\begin{equation}
ds^2 = N^2(t)dt^2 - a^2(t)\biggr[\frac{dr^2}{1 - kr^2} + r^2(d\theta^2 + \sin^2\theta d\phi^2)\biggl]\quad ,
\end{equation}
$N$ being the lapse function and $k = 1,0,-1$ correspond to a closed, flat and open
universe, respectively. Inserting this metric in the gravitational part of the Lagrangian
(\ref{Lag}), we obtain
\begin{equation}
L_G = N a^3\phi\biggr\{- \frac{6}{N^2}\biggr[\frac{\ddot a}{a} + \biggr(\frac{\dot a}{a}\biggl)^2
- \frac{\dot N}{N}\frac{\dot a}{a} + N^2\frac{k}{a^2}\biggl]
- \frac{\omega}{N^2}\biggr(\frac{\dot a}{a}\biggl)^2 - \frac{\phi^{n-1}}{N^2}\dot\chi^2\biggl\}
\quad .
\end{equation}
Integrating by part the second derivative of the scale factor and discarding surface terms,
we end up with the following expression for the gravitational Lagrangian:
\begin{equation}
L_G = 6\phi a\frac{\dot a^2}{N} + 6a^2\dot a\frac{\dot \phi}{N} - 6kNa\phi - \frac{\omega}{N}a^3\frac{\dot\phi^2}{\phi} - \phi^n\frac{a^3}{N}\dot\chi^2 
\quad .
\end{equation}
The presence of crossing terms between the scale factor and the dilatonic field is a
consequence of the non-minimal coupling in the original Lagrangian.
\par
Following the canonical procedure, the conjugate momenta are:
\begin{eqnarray}
\Pi_a &=& \frac{\partial L}{\partial\dot a} = 12a\dot a\frac{\phi}{N} + 6a^2\frac{\dot\phi}{N}
\quad ;\\
\Pi_\phi &=& \frac{\partial L}{\partial\dot\phi} = 6a^2\frac{\dot a}{N} - 2\omega\frac{a^3}{N}\frac{\dot\phi}{\phi} \quad ;\\
\Pi_\chi &=& \frac{\partial L}{\partial\dot\chi} = - 2\phi^n\frac{a^3}{N}\dot\chi \quad .
\end{eqnarray}
The final expression for the gravitational Hamiltonian is:
\begin{equation}
H_G = \frac{N}{3 + 2\omega}\biggr[\frac{\omega}{12}\frac{1}{a\phi}\Pi^2_a - \frac{1}{2}\frac{\phi}{a^3}\Pi^2_\phi + \frac{1}{2a^2}\Pi_a\Pi_\phi\biggl] -
\frac{1}{4}\phi^{-n}\frac{N}{a^3}\Pi^2_\chi + 6kNa\phi \quad .
\end{equation}
\par
The material part of the Lagrangian can be treated using the Schutz's formalism, which
has the advantage of attributing degrees of freedom to the matter field. In this formalism
this matter Lagrangian is written as
\begin{equation}
L_m = \sqrt{-g}p
\end{equation}
where $p$ is the pressure. We will consider a barotropic equation of state:
$p = \alpha\rho$, $\alpha \leq 1$.
The fluid's four-velocity is expressed in terms of five potentials $\epsilon$,
$\zeta$, $\beta$, $\theta$ and $S$:
\begin{equation}
U_\nu = \frac{1}{\mu}(\epsilon_{,\nu} + \zeta\beta_{,\nu} +
\theta S_{,\nu})
\end{equation}
where $\mu$ is the specific enthalpy. The variable $S$ is the specific
entropy, while the potentials $\zeta$ and $\beta$ are connected with
rotation and are absent for models of the Friedmann-Robertson-Walker (FRW) type. The variables $\epsilon$ and
$\theta$ have no clear physical meaning.
The four-velocity is subject to the normalization condition
\begin{equation}
U^\nu U_\nu = -1 \quad .
\end{equation}
Using the constraint for the fluids and after some thermodynamical considerations
\cite{schutz1,schutz2},
the matter Lagrangian takes the form:
\begin{equation}
L_m = N^{-1/\alpha}
a^3\frac{\alpha}{(\alpha + 1)^{1/\alpha + 1}}(\dot\epsilon +
\theta\dot S)^{1/\alpha + 1}\exp\biggr(- \frac{S}{\alpha}\biggl) \quad .
\end{equation}
This matter Lagrangian may be further simplified  leading,
by canonical methods \cite{rubakov}, to the
matter Hamiltonian
\begin{equation}
H_m =  p_\epsilon^{\alpha + 1} a^{-3\alpha}e^S
\end{equation}
where $p_\epsilon = -\rho_0 U^0 Na^3$,
$\rho_0$ being the rest mass density of the fluid.
The  canonical transformation
\begin{equation}
T = -p_Se^{-S}p_\epsilon^{-(\alpha + 1)} \quad , \quad
p_T = p_\epsilon^{\alpha + 1}e^S \quad , \quad
\bar\epsilon = \epsilon - (\alpha + 1)\frac{p_S}{p_\epsilon} \quad ,
\quad \bar p_\epsilon = p_\epsilon \quad ,
\end{equation}
which generalizes the one used in \cite{rubakov},
takes the  total Hamiltonian (gravitational plus matter) to the  final
form
\begin{equation}
H = \frac{N}{3 + 2\omega}\biggr[\frac{\omega}{12}\frac{1}{a\phi}\Pi^2_a - \frac{1}{2}\frac{\phi}{a^3}\Pi^2_\phi + \frac{1}{2a^2}\Pi_a\Pi_\phi\biggl] -
\frac{1}{4}\phi^{-n}\frac{N}{a^3}\Pi^2_\chi + 6kNa\phi
+ \frac{p_T}{a^{3\alpha}} \quad ,
\end{equation}
where the momentum
$p_T$ is the only remaining canonical variable associated with matter.
\par
Quantizing this system by canonical methods, which mounts up to replace the
momenta by operators and considering the constraint
\begin{equation}
\hat H\Psi = 0 \quad ,
\end{equation}
where $\Psi$ is the wave function of the Universe,
we obtain the Wheeler-DeWitt equation in the minisuperspace:
\begin{equation}
\label{wdw1}
\biggr\{\frac{N}{3 + 2\omega}\biggr[- \frac{\omega}{12}\frac{1}{a\phi}\partial^2_a + \frac{1}{2}\frac{\phi}{a^3}\partial^2_\phi - \frac{1}{2a^2}\partial_a\partial_\phi\biggl] +
\frac{1}{4}\phi^{-n}\frac{N}{a^3}\partial^2_\chi + 6kNa\phi
- i\frac{1}{a^{3\alpha}}\partial_T\biggl\}\Psi = 0 \quad .
\end{equation}
The canonical transformation employed for the matter fields fixes the parametrization
of the time $T$ \cite{nivaldo}. For pressurelless matter this parametrization corresponds to the cosmic time and
for the radiative fluid to the conformal time.
\par
The presence of cross derivatives makes the analysis of the Wheeler-DeWitt equation
(\ref{wdw1}) quite delicate. But, we can
diagonalize the Wheeler-DeWitt equation through the change of
variables
\begin{equation}
\label{change}
a \rightarrow \phi^{-1/2}b \quad , \quad \phi \rightarrow \phi \quad .
\end{equation}
In terms of these new variables, the Wheeler-DeWitt equation takes the form,
\begin{equation}
\label{wdw2}
\biggr\{- \frac{1}{24b}\partial^2_b + \frac{1}{4\tilde\omega}\frac{\phi^2}{b^3}\biggr[\partial^2_\phi + \frac{2}{\phi}\partial_\phi\biggl]
+ \frac{1}{4}\frac{\phi^{1-n}}{b^3}\partial^2_\chi + 6kb - i\frac{\phi^{(3\alpha-1)/2}}{b^{3\alpha}}\partial_T\biggl\}\Psi = 0 \quad ,
\end{equation}
where $\tilde\omega = \omega + 3/2$. The first derivative in the field $\phi$ is an
ordering term introduced to assure the hermiticity of the effective Hamiltonian.
The final results depend very weakly on this term. Remark that the change of variable (\ref{change}) corresponds
to perform a conformal transformation in the original Lagrangian (\ref{Lag}), passing from
the so-called Jordan's frame to the Einstein's frame \cite{nelson1}. In this sense, $\omega < - 3/2$ corresponds
to a scalar field with negative energy in the Einstein's frame. When $\omega = - 3/2$, the dilatonic field is null in the Einstein's frame.

\section{Wave functions and wave packets}

The task to be addressed now is to solve the equation (\ref{wdw2}). To do so, we consider
first stationary states given by
\begin{equation}
\Psi = \Phi e^{-iET} \quad .
\end{equation}
In order to determine the solutions of (\ref{wdw2}), the method of separation of variables is
employed. This implies to write 
\begin{equation}
\Phi(b,\phi,\chi) = X(b)Y(\phi)Z(\chi) \quad .
\end{equation}
The Wheeler-DeWitt equation reduces to
\begin{equation}
- \frac{1}{24b}\frac{X_{bb}}{X} + \frac{1}{4\tilde\omega}\frac{\phi^2}{b^3}\biggr[\frac{Y_{\phi\phi}}{Y} + \frac{2}{\phi}\frac{Y_\phi}{Y}\biggl] + \frac{1}{4}\frac{\phi^{1 - n}}{b^3}\frac{Z_{\chi\chi}}{Z}
+ 6kb = E\frac{\phi^{(3\alpha - 1)/2}}{b^{3\alpha}} \quad ,
\end{equation}
where the subscripts indicate derivative with respect to the variables $b$, $\phi$ and
$\chi$. The solution for the function $Z$ is
\begin{equation}
\frac{Z_{\chi\chi}}{Z} = - r^2 \quad \Rightarrow \quad Z = Z_0e^{\pm ir\chi}\quad , \quad
r \in \Re \quad .
\end{equation}
\par
In order to proceed, we particularize to the case $\alpha = 1/3$, that is, matter is a radiative
fluid. This case is of special importance because an eletromagnetic term is present in the
effective action from string theories in the Ramond-Ramond sector. We end with two equations:
\begin{eqnarray}
\label{eq1}
X'' +\biggr\{ -\frac{24}{b^2}s - 144kb^2 + E\biggl\}X &=& 0 \quad , \\
\label{eq2}
\ddot Y + 2\frac{\dot Y}{\phi} - \biggr\{r^2\tilde\omega\phi^{-(1+n)} + 4s\frac{\tilde\omega}{\phi^2}\biggl\}Y &=& 0 \quad ,
\end{eqnarray}
where $s$ is a separation constant which will play a crucial role in the analysis of
the quantum model. We have also made the redefinition $24E \rightarrow E$.
\par
Equations (\ref{eq1},\ref{eq2}) can be solved in terms of Bessel functions when
$k = 0$. The solution
depends on the sign of $\tilde\omega$. The final
solutions for the wave function in the spatial flat case are:
\begin{itemize}
\item $\tilde\omega > 0$ ($\omega > -3/2$):
\begin{equation}
\label{wf1}
\Psi(a,\phi,\chi,t) = A\sqrt{\frac{a}{\sqrt{\phi}}} J_\mu\biggr(\sqrt{E}a\phi^{1/2}\biggl)
K_\nu\biggr(r\frac{\sqrt{\tilde w}}{p}\phi^p\biggl)e^{-iEt \pm ir\chi} \quad ;
\end{equation}
\item $\tilde\omega < 0$ ($\omega < - 3/2$):
\begin{equation}
\label{wf2}
\Psi(a,\phi,\chi,t) = A\sqrt{\frac{a}{\sqrt{\phi}}} J_\mu\biggr(\sqrt{E}a\phi^{1/2}\biggl)
J_\nu\biggr(r\frac{\sqrt{|\tilde w}|}{p}\phi^p\biggl)e^{-iEt \pm ir\chi} \quad .
\end{equation}
\end{itemize}
In these expressions, $\mu = \sqrt{24s +1/4}$, $\nu = (1/2p)\sqrt{16\tilde\omega s +1}$, $p = (1 - n)/2$,
and $J_\nu$ and $K_\nu$ are the ordinary and modified Bessel functions.
By restricting the parameter $s$ to a convenient range of values, the wave functions
(\ref{wf1},\ref{wf2}) satisfies the required boundary conditions in order to guarantee the hermiticity
of the Hamiltonian, but, they are not in general square integrable. So, in order to assure
the consistency of the quantum model, a wave packet must be constructed. This can be
achieved by performing a gaussian integration on the separation parameters $E$, $r$ and $s$.
This must be done separately for each sign of $\tilde\omega$.
\par
For $\tilde\omega > 0$, let us consider the wave packet given by
\begin{eqnarray}
\Psi(a,\phi,\chi,T) &=& \int_0^\infty\int_0^\infty\int_{s_1}^{s_2}ds\,dx\,dy A(s)x^{\mu +1}e^{-\rho(T) x^2}y^{\mu - 1}e^{-\sigma(\chi) y}\sqrt{\frac{a}{\sqrt{\phi}}}\times\nonumber\\
&\times& J_\mu\biggr(x a\phi^{1/2}\biggl)
K_\nu\biggr(y\frac{\sqrt{\tilde w}}{p}\phi^p\biggl) \quad ,
\end{eqnarray}
where we have defined $x \equiv \sqrt{E}$, $y \equiv r$, $\rho(T) = a_1 + iT$, $\sigma(\chi) = a_2 + i\chi$. Moreover, $a_{1,2}$ are
positive constants and $A(s)$ is a function to be analyzed later.
Performing the integrations on $x$ and $y$ \cite{grad1}, we find
\begin{eqnarray}
\label{wp1}
\Psi(a,\phi,\chi,T) &=& \int_{s_1}^{s_2}ds\,A(s)\sqrt{\frac{a}{\sqrt{\phi}}}\frac{(a\phi^{1/2})^\mu}{(2\rho(T))^{\mu + 1}}
e^{- \frac{a^2\phi}{4\rho(T)}}\biggr(2\frac{\sqrt{\tilde\omega}}{p}\phi^p\biggl)^\nu
\frac{1}{\biggr[\sigma(\chi) + \frac{\sqrt{\tilde\omega}}{p}\phi^p\biggl]^{\mu + \nu}}
\times\nonumber\\
&\times&\frac{\Gamma(\mu + \nu)\Gamma(\mu - \nu)}{\Gamma(\mu + 1/2)}\,_2F_1\biggr(\mu + \nu, \nu + 1/2,\mu + 1/2, \frac{\sigma(\chi) - \frac{\sqrt{\tilde\omega}}{p}\phi^p}{\sigma(\chi) + \frac{\sqrt{\tilde\omega}}{p}\phi^p}\biggl) \quad,
\end{eqnarray}
where $_2F_1(\alpha,\beta,\gamma,z)$ is the hypergeometric function and some unimportant
constants were absorbed in the factor $A$.
\par
For $\tilde\omega < 0$, we choose, with the same definitions as in the
preceding case, the superposition given by
\begin{eqnarray}
\Psi(a,\phi,\chi,T) &=& \int_0^\infty\int_0^\infty\int_{s_1}^{s_2}ds\,dx\,dy A(s)x^{\mu +1}e^{-\rho(T) x^2}y^{\nu + 1}e^{-\sigma(\chi) y}\sqrt{\frac{a}{\sqrt{\phi}}}\times\nonumber\\
&\times& J_\mu\biggr(x a\phi^{1/2}\biggl)
J_\nu\biggr(y\frac{\sqrt{|\tilde w|}}{p}\phi^p\biggl) \quad ,
\end{eqnarray}
leading, after integration \cite{grad2}, to the wave function
\begin{eqnarray}
\Psi(a,\phi,\chi,T) &=& \int_{s_1}^{s_2}ds\,A(s)\sqrt{\frac{a}{\sqrt{\phi}}}\frac{(a\phi^{1/2})^\mu}{(2\rho(T))^{\mu + 1}}
e^{- \frac{a^2\phi}{4\rho(T)}}\sigma(\chi)\biggr(2\frac{\sqrt{|\tilde\omega|}}{p}\phi^p\biggl)^\nu
\times\nonumber\\
&\times&\frac{\Gamma(\nu + 3/2)}{\biggr[\sigma^2(\chi) + \frac{|\tilde\omega|}{p^2}\phi^{2p}\biggl]^{\nu + 3/2}}
\end{eqnarray}

\section{String quantum cosmological models}

In the perfect fluid employed above, a time coordinate may be identified with the matter
variables. Hence, a Schr\"odinger-like equation has been obtained. From the solutions
found, it is possible to determine a cosmological model through the computation of
the expectation values for the dynamical variables $a$, $\phi$ and $\chi$. This computation
follows the ordinary procedure of quantum mechanics in the Schr\"odinger picture:
\begin{equation}
<x(T)> = \frac{\int \Psi^*x\Psi d\Omega}{\int \Psi^*\Psi d\Omega} \quad ,
\end{equation}
where $d\Omega$ stands for the integration on the ensemble of variables.
But, in order this computation can make sense in the spirit of usual quantum mechanics,
the norm of the wave function must be time independent, otherwise there is no
conservation of probability, and the unitarity is lost. Moreover, this norm must be finite.
The norm of the wave function is given, as usual, by
\begin{equation}
N = \int_0^\infty\int_0^\infty\int_{-\infty}^\infty \Psi^*(a,\phi,\chi)\Psi(a,\phi,\chi)
\,da\,d\phi\,d\chi \quad .
\end{equation}
We have restricted the integration on $a$ and $\phi$ to the positive semi-real axis due
to physical grounds (anyway, negative values would give non-convergent results), while
the axionic field is allowed to take values in all real axis (in any case, the boundary
conditions require that the axionic field may also take negative values).
We turn now to the analysis of the norm of the wave function in the two kind of
situations we have: the "normal" case ($\tilde\omega > 0$) and the "anomalous" case
($\tilde\omega < 0$).

\subsection{The normal case ($\tilde\omega > 0$)}

The wave function is given by (\ref{wp1}). It presents some pathologies as we will now
discuss. Let us consider
\begin{eqnarray}
\Psi^*\Psi &=& \int_{s_1}^{s_2}\int_{s'_1}^{s'_2}\,ds\,ds'\,A(s)A(s')\frac{a}{\sqrt{\phi}}
\frac{(a\phi^{1/2})^{\mu + \mu'}}{(2\rho(T))^{\mu + 1}(2\rho^*(T))^{\mu' + 1}}
\exp\biggr[-\frac{a_1a^2\phi}{2\rho(T)\rho^*(T)}\biggl]\nonumber\\
&\times&\biggr(2\frac{\sqrt{\tilde\omega}}{p}\phi^p\biggl)^{\nu
+ \nu'}
\frac{1}{\biggr[\sigma(\chi) + \frac{\sqrt{\tilde\omega}}{p}\phi^p\biggl]^{\mu + \nu}}
\frac{1}{\biggr[\sigma^*(\chi) + \frac{\sqrt{\tilde\omega}}{p}\phi^p\biggl]^{\mu' + \nu'}}
\nonumber\\
&\times&\frac{\Gamma(\mu + \nu)\Gamma(\mu - \nu)\Gamma(\mu' + \nu')\Gamma(\mu' - \nu')}{
\Gamma(\mu + 1/2)\Gamma(\mu'+ 1/2)}\nonumber\\
&\times&_2F_1\biggr(\mu + \nu,\nu + 1/2,\mu + 1/2, \frac{\sigma(T) - \frac{\sqrt{\tilde\omega}}{p}\phi^p}{\sigma(T) + \frac{\sqrt{\tilde\omega}}{p}\phi^p}\biggl)\nonumber\\
&\times&_2F_1\biggr(\mu + \nu,\nu + 1/2,\mu + 1/2, \frac{\sigma^*(T) - \frac{\sqrt{\tilde\omega}}{p}\phi^p}{\sigma^*(T) + \frac{\sqrt{\tilde\omega}}{p}\phi^p}\biggl)
\end{eqnarray}
The integration in the variable $a$ can be easily performed, leading to
\begin{eqnarray}
\int_0^\infty\Psi^*\Psi\,da &=& \int_{s_1}^{s_2}\int_{s'_1}^{s'_2}\,ds\,ds'\,\frac{A(s)A(s')}{(2a_1)^{1 +\mu/2 + \mu'/2}}\frac{1}{\phi^{3/2}}
(2\rho(T))^{\mu - \mu'}(\rho^*(T))^{\mu' - \mu}\Gamma[1 + (\mu + \mu')/2]
\nonumber\\
&\times&\biggr(2\frac{\sqrt{\tilde\omega}}{p}\phi^p\biggl)^{\nu
+ \nu'}
\frac{1}{\biggr[\sigma(\chi) + \frac{\sqrt{\tilde\omega}}{p}\phi^p\biggl]^{\mu + \nu}}
\frac{1}{\biggr[\sigma^*(\chi) + \frac{\sqrt{\tilde\omega}}{p}\phi^p\biggl]^{\mu' + \nu'}}
\nonumber\\
&\times&\frac{\Gamma(\mu + \nu)\Gamma(\mu - \nu)\Gamma(\mu' + \nu')\Gamma(\mu' - \nu')}{
\Gamma(\mu + 1/2)\Gamma(\mu'+ 1/2)}\nonumber\\
&\times&_2F_1\biggr(\mu + \nu,\nu + 1/2,\mu + 1/2, \frac{\sigma(T) - \frac{\sqrt{\tilde\omega}}{p}\phi^p}{\sigma(T) + \frac{\sqrt{\tilde\omega}}{p}\phi^p}\biggl)\nonumber\\
&\times&_2F_1\biggr(\mu + \nu,\nu + 1/2,\mu + 1/2, \frac{\sigma^*(T) - \frac{\sqrt{\tilde\omega}}{p}\phi^p}{\sigma^*(T) + \frac{\sqrt{\tilde\omega}}{p}\phi^p}\biggl)
\end{eqnarray}
Writting $\rho(T) = Re^{i\Theta}$, where $R = \sqrt{a_1^2 + T^2}$, $\Theta = \arctan(T/a_1)$,
the final expression is
\begin{eqnarray}
\label{int1}
\int_0^\infty\Psi^*\Psi\,da &=& \int_{s_1}^{s_2}\int_{s'_1}^{s'_2}\,ds\,ds'\,\frac{A(s)A(s')}{(2a_1)^{1 +\mu/2 + \mu'/2}}\frac{1}{\phi^{3/2}}
\cos\biggr[(\mu - \mu')\Theta(T)\biggl]\Gamma[1 + (\mu + \mu')/2]
\nonumber\\
&\times&\biggr(2\frac{\sqrt{\tilde\omega}}{p}\phi^p\biggl)^{\nu
+ \nu'}
\frac{1}{\biggr[\sigma(\chi) + \frac{\sqrt{\tilde\omega}}{p}\phi^p\biggl]^{\mu + \nu}}
\frac{1}{\biggr[\sigma^*(\chi) + \frac{\sqrt{\tilde\omega}}{p}\phi^p\biggl]^{\mu' + \nu'}}
\nonumber\\
&\times&\frac{\Gamma(\mu + \nu)\Gamma(\mu - \nu)\Gamma(\mu' + \nu')\Gamma(\mu' - \nu')}{
\Gamma(\mu + 1/2)\Gamma(\mu'+ 1/2)}\nonumber\\
&\times&_2F_1\biggr(\mu + \nu,\nu + 1/2,\mu + 1/2, \frac{\sigma(T) - \frac{\sqrt{\tilde\omega}}{p}\phi^p}{\sigma(T) + \frac{\sqrt{\tilde\omega}}{p}\phi^p}\biggl)\nonumber\\
&\times&_2F_1\biggr(\mu + \nu,\nu + 1/2,\mu + 1/2, \frac{\sigma^*(T) - \frac{\sqrt{\tilde\omega}}{p}\phi^p}{\sigma^*(T) + \frac{\sqrt{\tilde\omega}}{p}\phi^p}\biggl)
\end{eqnarray}
\par
The crucial point is that the expression (\ref{int1}) is time-dependent unless
$s = s'$. Further integrations on $\phi$ and $\chi$ can not change this situation.
Hence, in order to assure the unitarity of the quantum model we must set
\begin{equation}
A(s) = \delta(s - s_0) \quad ,
\end{equation}
where $s_0$ is an arbitrary number in the range of values of $s$ which lead to the correct
boundary condition. Hence, we obtain
\begin{eqnarray}
\label{int2}
N &=& \int_0^\infty\int_0^\infty\int_{-\infty}^{\infty}\Psi^*\Psi\,da\,d\phi\,d\chi = \int_0^\infty
\int_{-\infty}^\infty d\phi\,d\chi \frac{1}{(2a_1)^{1 +\mu}}\frac{1}{\phi^{3/2}}
\biggr(2\frac{\sqrt{\tilde\omega}}{p}\phi^p\biggl)^{2\nu}\Gamma(1 + \mu)\nonumber\\
&\times&\frac{1}{\biggr\{\biggr[\sigma(\chi) + \frac{\sqrt{\tilde\omega}}{p}\phi^p\biggl]
\biggr[\sigma^*(\chi) + \frac{\sqrt{\tilde\omega}}{p}\phi^p\biggl]\biggl\}^{\mu + \nu}}
\frac{(\Gamma(\mu + \nu)\Gamma(\mu - \nu))^2}{\Gamma^2(\mu + 1/2)}\nonumber\\
&\times&_2F_1\biggr(\mu + \nu,\nu + 1/2,\mu + 1/2, \frac{\sigma(T) - \frac{\sqrt{\tilde\omega}}{p}\phi^p}{\sigma(T) + \frac{\sqrt{\tilde\omega}}{p}\phi^p}\biggl) \nonumber\\
&\times&_2F_1\biggr(\mu + \nu,\nu + 1/2,\mu + 1/2, \frac{\sigma^*(T) - \frac{\sqrt{\tilde\omega}}{p}\phi^p}{\sigma^*(T) + \frac{\sqrt{\tilde\omega}}{p}\phi^p}\biggl)
\quad .
\end{eqnarray}
The norm is time independent. Now, in principle we could proceed and compute the expectation value for
the dynamical variables. However, the norm of the wave function is divergent. The construction
of the wave packet has not erased the divergences present in the norm of the eigenmodes. Even
if we can not assure that this divergence appears for all wave packets, it is suggestive that
others wave packets that can be constructed using more complicated expressions, 
disposable in integral tables, lead to the same problem.
\par
Since the norm of the wave function is divergent, the only possible consistent
situation is to fix
the gravitational coupling constant from the begining. In this case, we return back to
General Relativity. Moreover, since $\phi$ is fixed from the begining, it is a free parameter.
No prediction concerning the dilaton field can be obtained in this case.

\subsection{The anomalous case ($\tilde\omega < 0$)}

We compute again the norm of the wave function. It is given by
\begin{eqnarray}
N &=& \int_0^\infty\int_0^\infty\int_{-\infty}^\infty\Psi^*(a,\phi,\chi)\Psi(a,\phi,\chi)\,da\,d\phi\,d\chi \nonumber\\
&=& \int_0^\infty\int_0^\infty\int_{-\infty}^\infty\int_{s_1}^{s_2}\int_{s'_1}^{s'_2}\,A(s')\,A(s)\frac{a}{\sqrt{\phi}}
\frac{(a\phi^{1/2})^{\mu + \mu'}}{(2\rho(T))^{\mu + 1}(2\rho^*(T))^{\mu' + 1}}\nonumber\\
&\times&
\exp\biggr[- \frac{a_1a^2\phi}{2\rho(T)\rho^*(T)}\biggl]\sigma(\chi)\sigma^*(\chi)\biggr(2\frac{\sqrt{|\tilde\omega|}}{p}\phi^p\biggl)^{\nu + \nu'}
\nonumber\\
&\times&\frac{\Gamma(\nu + 3/2)\Gamma(\nu' + 3/2)}{\biggr[\sigma^2(\chi) + \frac{|\tilde\omega|}{p^2}\phi^{2p}\biggl]^{\nu + 3/2}\biggr[(\sigma^*)^2(\chi) + \frac{|\tilde\omega|}{p^2}\phi^{2p}\biggl]^{\nu' + 3/2}}\,ds\,ds'\,da\,d\phi\,d\chi
\end{eqnarray}
The integration on $a$ can be performed, leading to
\begin{eqnarray}
N &=& \int_0^\infty\int_{-\infty}^\infty\int_{s_1}^{s_2}\int_{s'_1}^{s'_2}\,A(s')\,A(s)\frac{1}{\sqrt{\phi^3}}
\frac{\rho(T)^{(\mu'- \mu)/2}\rho^*(T)^{(\mu - \mu')/2}}{(2a_1)^{1 + \mu/2 + \mu'/2}}
\nonumber\\
&\times&
\sigma(\chi)\sigma^*(\chi)\biggr(2\frac{\sqrt{|\tilde\omega|}}{p}\phi^p\biggl)^{\nu + \nu'}
\nonumber\\
&\times&\frac{\Gamma[1 + (\mu + \mu')/2]\Gamma(\nu + 3/2)\Gamma(\nu' + 3/2)}{\biggr[\sigma^2(\chi) + \frac{|\tilde\omega|}{p^2}\phi^{2p}\biggl]^{\nu + 3/2}\biggr[(\sigma^*)^2(\chi) + \frac{|\tilde\omega|}{p^2}\phi^{2p}\biggl]^{\nu' + 3/2}}\,ds\,ds'\,d\phi\,d\chi
\end{eqnarray}
Writting, as before, $\rho(T) = Re^{i\Theta}$, with $R(T) = \sqrt{a_1^2 + T^2}$, $\Theta(T) = \arctan(T/a_1)$,
and using the fact that, aside these terms, the rest of the integrand is symmetric by
the interchange $s \leftrightarrow s'$, we end up with the integral expression
\begin{eqnarray}
N &=& \int_0^\infty\int_{-\infty}^\infty\int_{s_1}^{s_2}\int_{s'_1}^{s'_2}\,A(s')\,A(s)\frac{1}{\sqrt{\phi^3}}
\frac{\cos[(\mu'- \mu)\Theta(T)]}{(2a_1)^{1 + \mu/2 + \mu'/2}}
\nonumber\\
&\times&
\sigma(\chi)\sigma^*(\chi)\biggr(2\frac{\sqrt{|\tilde\omega|}}{p}\phi^p\biggl)^{\nu + \nu'}
\nonumber\\
&\times&\frac{\Gamma[1 + (\mu + \mu')/2]\Gamma(\nu + 3/2)\Gamma(\nu' + 3/2)}{\biggr[\sigma^2(\chi) + \frac{|\tilde\omega|}{p^2}\phi^{2p}\biggl]^{\nu + 3/2}\biggr[(\sigma^*)^2(\chi) + \frac{|\tilde\omega|}{p^2}\phi^{2p}\biggl]^{\nu' + 3/2}}\,ds\,ds'\,d\phi\,d\chi
\end{eqnarray}
The norm of the wave function is time dependent unless
$A(s) = \delta(s - s_0)$ and $A(s') = \delta(s'- s_0)$.
Hence,
\begin{eqnarray}
N &=&  \int_0^\infty\int_{-\infty}^\infty\frac{1}{\sqrt{\phi^3}}
\frac{1}{(2a_1)^{1 + \mu}}
\sigma(\chi)\sigma^*(\chi)\biggr(2\frac{\sqrt{|\tilde\omega|}}{p}\phi^p\biggl)^{2\nu}
\nonumber\\
&\times&\frac{\Gamma(1 + \mu)\Gamma^2(\nu + 3/2)}{\biggr\{\biggr[\sigma^2(\chi) + \frac{|\tilde\omega|}{p^2}\phi^{2p}\biggl]\biggr[(\sigma^*(\chi))^2 + \frac{|\tilde\omega|}{p^2}\phi^{2p}\biggl]\biggl\}^{\nu + 3/2}}\,d\phi\,d\chi
\end{eqnarray}
The norm now is finite. 
\par
We proceed evaluating the expectation value of the wave function.
\par
The expectation value for a dynamical variable $X$ is given by
\begin{equation}
<X> = \frac{1}{N}\int_0^\infty\int_0^\infty\int_{-\infty}^{+\infty}\Psi^*X\Psi\,da\,d\phi\,d\chi
\quad .
\end{equation}
In general, it is possible to perform the integration in $a$, but the integration
in the variables $\phi$ and $\chi$ can only be made numerically.
\par
The result for the scale factor is:
\begin{eqnarray}
\label{exp-1}
<a> &=& \sqrt{a_1^2 + T^2}\,\,\frac{2^{2\nu + 1}}{pN}\,\frac{1}{(2a_1)^{\mu + 3/2}}
\biggr(\frac{p}{\sqrt{|\tilde\omega|}}\biggl)^{-\frac{1}{p}}\int_0^\infty
\int_{-\infty}^{+\infty}\,dz\,d\chi\,z^{- \frac{1}{p} + 1 + 2\nu}\nonumber\\
&\times&\Gamma^2(\nu + 3/2)
\Gamma(\mu + 3/2)(a_2^2 + \chi^2)\frac{1}{\biggr[(a_2^2 + z^2 - \chi^2)^2 + 4a_2^2\chi^2\biggl]^{\nu + 3/2}}
\quad ,
\end{eqnarray}
where $z = |\tilde\omega|\phi^p$. The time behavior is given by the first term in the
right hand side of (\ref{exp-1}): the scale factor exhibits a bounce and asymptotically
it behaves like the FRW radiation-dominated model (remember that for the radiative
fluid $T$ is the conformal time).
\par
Performing the same calculation for the dilatonic field, we find
\begin{eqnarray}
\label{exp-2}
<\phi> &=& \frac{2^{2\nu + 1}}{pN}\,\frac{1}{(2a_1)^{\mu + 3/2}}
\biggr(\frac{p}{\sqrt{|\tilde\omega|}}\biggl)^{\frac{1}{p}}\int_0^\infty
\int_{-\infty}^{+\infty}\,dz\,d\chi\, z^{\frac{1}{2p} - 1 + 2\nu}\nonumber\\
&\times&\Gamma^2(\nu + 3/2)
\Gamma(\mu + 1)(a_2^2 + \chi^2)\frac{1}{\biggr[(a_2^2 + z^2 - \chi^2)^2 + 4a_2^2\chi^2\biggl]^{\nu + 3/2}}
\quad .
\end{eqnarray}
Remark that the expectation value of $\phi$ does not depend on time. Finally,
it is easy to see that
\begin{equation}
<\chi> = 0 \quad ,
\end{equation}
since the integrand is a odd function and the integration is performed in
all real axis.
\par
The expectation value of the dilatonic is time independent, but we can obtain some predictions.
We have worked in units such that $\hbar = c = 1$. In this case, the Planck Mass is
the inverse of the Planck length. Let us consider the situation where the scale factor
carries a dimension of length. The dilatonic field has the dimension of $M_{Pl}^{2}$, since
it is connected with the inverse of the gravitational coupling. 
Hence, we can construct the dimensionless quantity
\begin{equation}
<a>_{min}\sqrt{<\phi>} = b
\end{equation}
where $b$ is a pure number and $<a>_{min}$ is the mininum value of the scale factor during
the bounce. Performing the numerical integration in the expressions for the scale factor,
we find that the minimum of the scale factor is of the order of $200\,L_{Pl}$ for $\omega =
- 1.6$ ($\tilde\omega = - 0.1$), $2\,L_{Pl}$ for $\omega = - 3$ ($\tilde\omega = - 1.5$) and
$10^{-5}\,L_{Pl}$ for $\omega = - 500$ ( $\tilde\omega$ essentially of the same order).
These results indicate that for values of $\omega$ near the critical value $- 3/2$, the minimum
value of the scale factor is far above the Planck length. As the value of $\omega$ grows in
absolute value, the minimum scale factor becomes highly transplanckian, and in the limit
$\omega \rightarrow - \infty$ the curvature singularity appears.

\subsection{Brans-Dicke quantum cosmological models}

The Brans-Dicke theory is a particular case of the general action (\ref{Lag}) when
$\chi =$ constant. Considering the wave functions previous settled out, the wave functions
for the Brans-Dicke case can be obtained as the limit $r \rightarrow 0$. Hence,
the wave function for the Brans-Dicke theory coupled to perfect fluid matter is
\begin{equation}
\label{bd}
\Psi(a,\phi,T) = \sqrt{\frac{a}{\sqrt{\phi}}}J_\mu(\sqrt{E}a\phi^{1/2})\biggr\{A\phi^{-p\nu}
+ B\phi^{p\nu}\biggl\}e^{-iT} \quad .
\end{equation}
The regularity of the wave function at the origin and at infinity implies
\begin{eqnarray}
\phi &\rightarrow& 0 \quad \Rightarrow \quad - \frac{1}{4} + \frac{\mu}{2} \pm p\nu > 0 \quad ;\\
\phi &\rightarrow& \infty \quad \Rightarrow \quad \pm p\nu < 0 \quad .
\end{eqnarray}
It is possible to obtain a range of the parameter $s$ where these conditions are
satisfied.
However, as in the normal string case discussed before, the wave functions 
(\ref{bd}) are not square integrable. In the computation of the norm of the wave functions
we find
\begin{eqnarray}
N &=& \int_0^\infty\int_0^\infty\,\Psi^*\Psi\,da\,d\phi
\nonumber\\
&=&
\int_0^\infty\int_0^\infty \phi^{-3/2}u\,J_\mu(ru)\,J_\mu(r'u)e^{-i(r^2 - r'^2)}\,du\,d\phi
\nonumber\\
&=&
\int_0^\infty\phi^{-3/2}\frac{1}{r}\delta(r - r')e^{-i(r^2 - r'^2)T}\,d\phi \quad ,
\end{eqnarray}
with $r = \sqrt{E}$ and $u = a\phi^{1/2}$.
There is a divergence when $\phi \rightarrow 0$.
\par
The divergence pointed out occurs for any sign of $\tilde\omega$. The consequence is that
no quantum cosmological model can be constructed in the pure Brans-Dicke case. As before, the
superposition of the wave functions through the construction of the wave packet, does not
alter this conclusion since we are only superposing functions that are not square integrable.
The only case that makes sense is $\phi =$ constant from the begining, that is,
the gravitational coupling has no dynamics at all. This leads to the general relativity quantum
cosmological model with $G \sim 1/\phi$. As in the normal string case, the value of
$G$ can not be determined and it becomes a free parameter of the model.

\section{Conclusions}

In this work, we have studied the quantum cosmological models
from the string effective action in
four dimensions coupled to a radiative fluid. The radiative have been described with the aid of
the Schutz's formalism. The quantization of this system in the minisuperspace have led
to a Schr\"odinger-like equation where the matter variables play the role of time. 
Since we ended up with a genuine Schr\"odinger equation the predictions for the evolution of the dynamical variables
were obtained through the computation of the corresponding expectation values.
\par
The result is somehow surprising: the dilatonic field must have a value independent of time
while the axionic field must be zero.
However, the reason for this result is different depending if we are treating the
normal case where $\omega > - 3/2$ or the anomalous case where $\omega < - 3/2$.
In the normal case, the wave functions obtained from the Schr\"odinger-like equation
are not square integrable. They have not led to square integrable expressions through the
construction of wave packets using analytical expressions - others wave packets besides that shown here in the text have been analyzed, but they exhibit the same problems. Hence, it seems that no
quantum cosmological model can be constructed unless the dilatonic field is fixed as a constant from the
begining, what reduces the string action to the general relativity action. In the anomalous
case, square integrable wave functions (consequently, wave packets also) can be obtained,
but the expectation value of the dilatonic field leads to a time independent expression,
while the expectation value for the axionic field is zero.
\par
The fact that the imposition to have square integrables functions leads to restriction on
the parameters of the quantum model is well known in ordinary quantum mechanics. In the
harmonic oscillator problem the quantization of the energy results from this imposition.
Here, if we want to be strict, accepting the string effective action (\ref{Lag}), with dynamical
dilatonic and axion fields, the
prediction we obtain is that $\omega < - 3/2$ and $<\phi>$ = constant. This excludes the
pure string case, $\omega = -1$. This exclusion, very strong of course, can be alleviate
by considering cases where no dynamics is attributed to the dilatonic field from the
begining. To consider the pure dilatonic model without the axion field does not
change this situation.
\par
The strength of any conclusion is directly related to the assumptions made. Here, one
important assumption is the presence of a genuine time coordinate associated to the
matter field. In other situations in quantum cosmology, such assumption has already led
to very curious result: if applied to a Bianchi type I cosmological model \cite{brasil2},
the norm of the wave packet is necessarily time dependent. One way to obtain some dynamics to the dilatonic field is to allow
the norm of the wave packet to depend on time here as in the
Bianchi type I case. However, as in that case, this leads to a non unitary quantum
model, what is conceptually a problem, unless some techniques coming from quantum open
system may be succesfully employed \cite{davies}. However, this lies outside the scope of the present work.
\newline
\vspace{0.5cm}
\newline
\noindent
{\bf Acknowledgements:} We thank CNPq (Brazil) for partial financial support.

\end{document}